\documentclass[aps,pre,showpacs,preprint,superscriptaddress]{revtex4}
\usepackage{graphicx}
\usepackage{subfigure}
\usepackage{amsmath}
\usepackage{ulem}
\usepackage{bm}
\usepackage{amssymb}
\usepackage{color}
\usepackage{epstopdf}
\usepackage{epsfig,dsfont,multirow}
\pdfoutput=1
\begin{document}
\title{Generation of bright collimated vortex $\gamma$-ray via laser driven cone-fan target}
\author{Cui-Wen Zhang}
\affiliation{Key Laboratory of Beam Technology of the Ministry of Education, and College of Nuclear Science and Technology, Beijing Normal University, Beijing 100875, China}
\author{Mamat-Ali Bake}
\affiliation{School of Physics Science and Technology, Xinjiang University, Urumqi 830046, China}
\author{Hong Xiao}
\affiliation{Key Laboratory of Beam Technology of the Ministry of Education, and College of Nuclear Science and Technology, Beijing Normal University, Beijing 100875, China}
\author{Hai-Bo Sang}
\affiliation{Department of Physics, Beijing Normal University, Beijing 100875, China}
\author{Bai-Song Xie \footnote{bsxie@bnu.edu.cn}}
\affiliation{Key Laboratory of Beam Technology of the Ministry of Education, and College of Nuclear Science and Technology, Beijing Normal University, Beijing 100875, China}
\affiliation{Institute of Radiation Technology, Beijing Academy of Science and Technology, Beijing 100875, China}
\date{\today}
\begin{abstract}
We use numerical simulations to demonstrate that a source of bright collimated vortex $\gamma$-ray with large orbital angular momentum can be achieved by irradiating a circularly polarized laser with an intensity about $10^{22}\rm{W/{cm^2}}$ on a cone-fan target. In the studied setup, electron beam of energy of hundreds of MeV and vortex laser pulse are formed. And furthermore a high quality vortex $\gamma$-ray is yielded with small divergence of $5^{\circ}$ and high peak brilliance $\sim5\times10^{22}$ photons ${\rm\cdot s^{-1} \cdot mm^{-2} \cdot mrad^{-2}}$ $0.1\%\mathrm{BW}$ at $10\mathrm{MeV}$. A considerable fraction of angular momentum of laser is converted to electron beam and vortex $\gamma$-ray, which are roughly $27.8\%$ and $3\%$, respectively. And the conversion efficiency of energy from laser to electron beam and vortex $\gamma$-ray are around $41\%$ and $3.8\%$. Moreover, comparative simulations for different right radius of cone reveal that there exists an optimal size that makes the highest angular momentum of $\gamma$-ray photons to be around $2.8\times10^6\hbar$. The comparative simulations for different laser modes exhibit that it is more appropriate to choose the circularly polarized laser to generate vortex $\gamma$-ray than the Laguerre-Gaussian one.
\end{abstract}
\maketitle

\section{INTRODUCTION}

The prospect of generating vortex $\gamma$-rays with orbital angular momentum (OAM) has recently attracted particular interest due to its potential applications in imaging \cite{Jesacher:2005,2}, material science \cite{3} and astrophysics \cite{4,5}. Many vortex $\gamma$-rays generation schemes involving circularly polarized (CP) lasers and/or Laguerre-Gaussian (LG) lasers have been proposed \cite{7,8,9,10,11,12,13,14,15,16,17,channel2021HPL,channel2022PST,cone-foil2018APL,2018NJP}. As we know, the CP laser carries only spin angular momentum (SAM) of $n\delta\hbar$, where $n$ is the laser photon number, $\delta$ is the SAM quantum number of a laser photon ($\delta=\pm1$ for the right and left-hand) and $\hbar$ is the reduced Planck constant. While the LG laser as a kind of vortex light, it has a helical wave front, a twisted Poynting vector and hollow transverse field distribution \cite{6}, thus, it carries both of OAM and SAM as $n(l+\delta)\hbar$, where $l$ represents the OAM quantum number of a laser photon, e.g., $l=0, \pm1, \pm2, \ldots$. Therefore, under the same intensity, the LG laser would carry more angular momentum (AM), i.e., the sum of OAM and SAM, than the CP laser. However, in a view point of practical experiment, the highest intensity of LG laser achieved at present is around ${\rm 6.3\times10^{19} W/cm^2}$ \cite{PRL2020LG}, which is much less than that of the CP laser ${\rm10^{23}W/cm^2}$ \cite{optica2021CP}, thus, this strongly restricts the application of the LG lasers for vortex $\gamma$-rays generation and related applications.

Vortex $\gamma$-rays can be generated by synchrotron radiation \cite{7,8} and nonlinear Compton scattering (NCS) \cite{9,10,11,12,13,14,15,16,17,channel2021HPL,channel2022PST,cone-foil2018APL,2018NJP} processes in quantum electrodynamics (QED) regime. The AM of laser can be transferred to the OAM of electrons first, and then be further transferred to the OAM of vortex $\gamma$-rays through synchrotron radiation. The AM conversion efficiency is $1.43\%$ \cite{7} and a vortex $\gamma$-ray with small divergence angle of $10^{\circ}$ has been obtained \cite{8}. In NCS, strong laser interacts with counter-propagating energetic electron beam got by laser wakefield acceleration \cite{2018MRE} or laser direct acceleration \cite{2020MRE}, and vortex $\gamma$-rays can be generated via $e^- +n\omega_0\rightarrow e^- + \omega_{\gamma}$, where $n>1$ is an integer, $\omega_0$ and $\omega_{\gamma}$ is the frequency of laser and vortex $\gamma$-ray, respectively. In Ref. \cite{10} Wang \textit{et al.} got a vortex $\gamma$-ray with OAM of $2.5\times10^{18}\hbar$ and energy up to hundreds of $\rm{MeV}$. Chen \textit{et al.} \cite{16} pointed out that the AM of the absorbed laser photons is not solely transferred to the emitted $\gamma$-ray photons, but due to radiation reaction shared between the $\gamma$-ray photons and interacting electrons. In the recent research Ref.\cite{channel2022PST}. it is found that the AM conversion efficiency is $2.1\%$. Therefore, NCS is a promising way for vortex $\gamma$-rays generation based on currently available laser and accelerator technologies. However, the enhancement of OAM, AM conversion efficiency, energy and collimation degree of vortex $\gamma$-rays is still challenging and need to be study furthermore.

To keep the advantage of CP laser intensity while overcome the deficiency of AM, fortunately, previous studies revealed that light-fan, i.e., spiral-shaped foil plasma \cite{PRL2014-fan,channel2021HPL} is a useful plasma mirror \cite{mirror2004PRE,mirror2012NP,mirror2013PRL} which can reflect the incident CP laser and transfer it into a vortex one by adding phase $\exp(il\phi)$.
Compare to tube targets \cite{11,channel2019PPCF,channel2021HPL,channel2022PST}, the cone targets can restrain laser better and the laser intensity can be increased up to several tens of times, the laser oscillating field can pull out the electrons in cone surface periodically and accelerate them forward via laser pondermotive force and result in overdense energetic electron beams \cite{cone2004POP,cone2012POP,cone2015POP,cone2016JAP}. Note that the AM conversion efficiency from laser to vortex $\gamma$-ray is only $0.67\%$ in the scheme by using tube target \cite{channel2021HPL}. And the vortex $\gamma$-ray in Ref.\cite{cone-foil2018APL} is found concentrating at $1\mathrm{MeV}$ when a CP-LG laser pulse is irradiating on a cone-foil target. Thus, the cone-fan target should be considered in vortex $\gamma$-ray generation.

Motivated by the cone-fan target advantages in both of increasing the laser intensity, the produced electron energy and the high AM of reflected laser, in this paper, we use three-dimensional (3D) particle-in-cell (PIC) simulations to perform numerical study of vortex $\gamma$-ray generation via CP laser irradiates a cone-fan target, note that both of the laser and fan are left-hand in our study. The electrons on the inner surface of cone can be extracted into cone-channel and then accelerated to the high energy. Moreover, the cone-channel can guide these electrons effectively and result in overdense energetic electron beam finally. Meanwhile, the SAM of CP laser is efficiently converted into the OAM of electrons. When the incident laser reaches the fan that attached on the rear side of cone, the laser is reflected and a vortex one is formed. Then, the latter collides head-on with energetic electron beam and a bright collimated vortex $\gamma$-ray is emitted through NCS. The OAM of vortex $\gamma$-ray comes from both the vortex laser and electrons.

During the whole process, the fan is used as a reflecting mirror to get vortex laser and spontaneously trigger the subsequent NCS. The cut-off energy and OAM of vortex $\gamma$-ray are up to $200\mathrm{MeV}$ and $5.9\times10^{18}\hbar$ respectively. Remarkably, the conversion rate of laser AM for electron beam and vortex $\gamma$-ray are around $27.8\%$ and $3\%$, and roughly $41\%$ of the laser energy is converted to electron beam and $3.8\%$ to vortex $\gamma$-ray. The peak brilliance of vortex $\gamma$-ray is about $5\times10^{22}$ photons ${\rm\cdot s^{-1} \cdot mm^{-2} \cdot mrad^{-2}}$ $0.1\%$ bandwidth at $10\mathrm{MeV}$ and divergence angle $5^{\circ}$. The properties of this bright collimated vortex $\gamma$-ray might enable the development of novel application in various domains.
\begin{figure}[htbp]\suppressfloats
\includegraphics[scale=0.55]{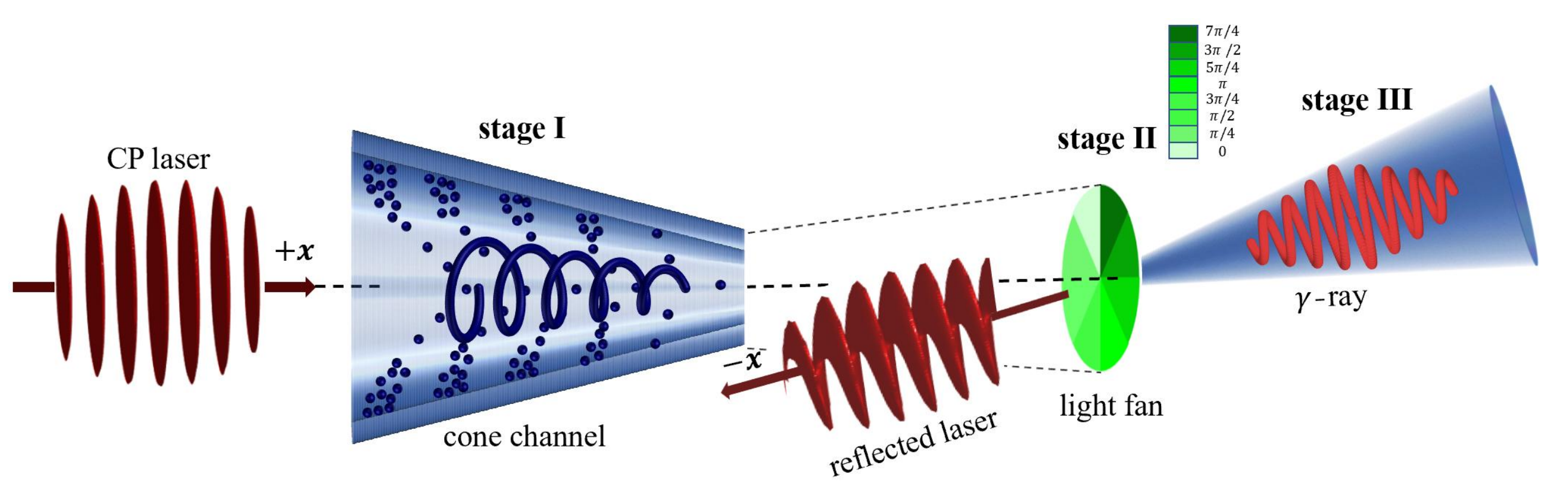}
\caption{\label{fig1}(color online).  The setup of the vortex $\gamma$-ray generation scheme. A CP laser  propagates forward along the $x$ axis from the left and irradiates a cone-fan target. The electrons on the inner surface of the cone can be extracted into the cone-channel and accelerated via direct laser acceleration, and overdense energetic electron beam can be generated. The incident laser is reflected when it reaches the fan and a vortex one is formed. Then, the latter collides head-on with the energetic electron beam and bright collimated vortex $\gamma$-ray is emitted through NCS.}
\end{figure}

\section{SCHEME SETUP AND SIMULATION PARAMETER}

\begin{figure}[htbp]\suppressfloats
\includegraphics[scale=0.65]{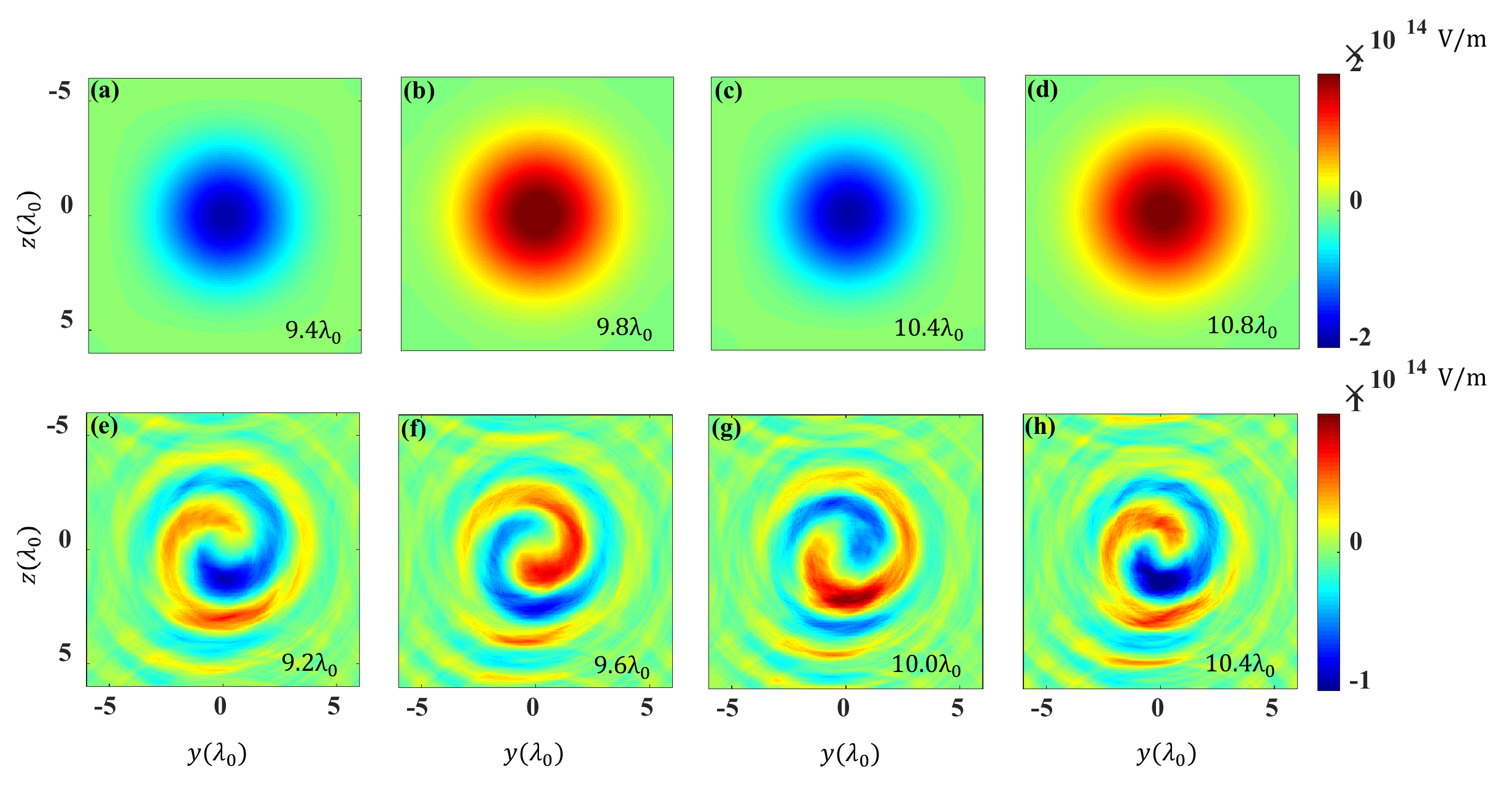}
\caption{\label{fig2}(color online). (a)-(d): Distributions of transverse electric field $E_y$ of incident CP laser at different crossing sections at $t=15T_0$. (e)-(h): Distributions of transverse electric field $E_y$ at different crossing sections at $t=25T_0$ when the incident laser is completely reflected by fan.}
\end{figure}

We performed 3D PIC simulations with open-source code EPOCH \cite{2014epoch,2015epoch}. As indicated in Fig.\ref{fig1}, a left-hand Gaussian CP laser propagates forward along the axis ($+x$ direction) of cone with a left-hand fan attached behind. The dimensionless amplitude of CP laser field is $a=a_0\exp(-(x/c-t_0))^2/\tau^2)\exp({-(y^2+z^2)/w^2})\left(\cos \varphi \widehat{\boldsymbol{e}}_{y}+\sin \varphi \widehat{\boldsymbol{e}}_{z}\right)$, where $a_0=eE_0/m_ec\omega_0=100$ (corresponding to $I_0\approx2.74\times 10^{22}\rm{W/{cm^2}}$) , $w=3\lambda_0$, $\lambda_0=cT_0=1\mathrm{\mu m}$ and $\omega_0=2\pi/T_0$ are the intensity, focal spot, wavelength and frequency of laser, respectively. $-e$ and $m_e$ are the electron charge and mass, $E_0$ is the electric field amplitude, $c$ is the speed of light in vacuum, $\varphi$ is the laser phase term and $\tau =10T_0$ is the laser pulse duration. The axial length of cone is $15\lambda_o$ located from $x=1\lambda_0$ to $x=16\lambda_0$. The inner left radius of cone $r_1=3\rm{\mu m}$, the right radius $r_2=1.5\rm{\mu m}$ and the thickness is $1\lambda_0$. The fan consists of eight parts with the same step height $\Delta=\lambda_0/16$ to mimic a $\lambda_0/2$ spiral phase plate with thickness increasing with angle and the minimum is $1\lambda_0$ to make sure that it can reflect the driving laser, and the right end of the fan is located at $x=16\lambda_0$. The grid size of simulation box is $40\lambda_0\times 12\lambda_0\times 12\lambda_0$ with $2000\times300\times300$ cells in the $x\times y\times z$ direction and there are 9 macro-particles in each cell. The density of electrons is $200n_c$, where $n_c=m_e\omega_0^2/4\pi e^2=1.1\times10^{21}\rm{cm^{-3}}$ is the critical density.

 In the first stage, shown in Fig.\ref{fig1}, electrons within the inner surface of cone are extracted into cone-channel periodically by oscillating laser field and are accelerated forward via laser pondermotive force. Due to the guiding action of cone, an overdense electron beam with energy of hundreds of $\mathrm{MeV}$ is generated. During the process, the SAM of CP laser is efficiently converted into the OAM of electrons, and we will discuss this point later.

In the second stage, the incident CP laser reaches fan and is reflected into a vortex one, see Fig.\ref{fig1}. To illustrate this process, as shown in Fig.\ref{fig2}, a simulation without cone is performed. Figures \ref{fig2}(e)-\ref{fig2}(h) indicate that the reflected laser pulse is in vortex shape, and due to the lost in the reflection process, the electric field of which is reduced to half that of the incident laser. The laser mode of reflected pulse can be expanded by a series of LG modes \cite{LGmode} and the amplitude of $\rm{LG_{nm}}$ can be defined by
\begin{equation}\label{1}
\begin{aligned}
\rm{LG_{nm}}(r,\phi,x)=&(C_{nm}/w)\exp(-ikr^2/2R)\exp(-r^2/w^2)\\
&\times \exp[-i(n+m+1)\psi]\\
&\times \exp[-i(n-m)\phi](-1)^p\\
&\times (r\sqrt2/w)^{|l|}L_p^{|l|}(2r^2/w^2),
\end{aligned}
\end{equation}
where
\begin{equation}\label{2}
\begin{aligned}
R(x)=(x_R^2+x^2)/x,
\end{aligned}
\end{equation}

\begin{equation}\label{3}
\begin{aligned}
w(x)=[2(x_R^2+x^2)/kx_R]^{1/2},
\end{aligned}
\end{equation}

\begin{equation}\label{4}
\begin{aligned}
\psi(x)=\arctan(x/x_R),
\end{aligned}
\end{equation}
$r$, $\phi$ and $x$ are cylindrical coordinates, $C_{nm}$ is the normalization constant, $k=2\pi/\lambda$ is the wave number, $x_R$ is the Rayleigh range, ${L_p}^l(2r^2/w^2)$ is the generalized Laguerre polynomial,  $l=|n-m|$ is the azimuthal mode index, and $p=\min(n,m)$ is the number of radial nodes. The Gaussian mode of incident laser beam in simulation is $\rm{LG_{00}}$ whose wave front has been modified by fan structure and the mode decomposition can be written using expansion coefficients \cite{LGmode}
\begin{equation}\label{5}
\begin{aligned}
a_{s t}=&\left\langle\mathrm{LG}_{s t}|\exp (-i \Delta \phi)| \mathrm{LG}_{nm}\right\rangle \\
=& \iint r d r d \phi\left(C_{s t}^{*} / w_{s t}\right) \exp \left(i k r^{2} / 2 R_{s t}-r^{2} / w_{s t}^{2}\right) \\
& \times \exp [-i(s-t) \phi](-1)^{\min (s, t)}\left(r \sqrt{2} / w_{s t}\right)^{|s-t|} \\
& \times L_{\min (s, t)}^{|s-t|}\left(2 r^{2} / w_{s t}^{2}\right) \exp (-i \Delta \phi) \\
& \times\left(C_{00} / w\right) \exp \left(-i k r^{2} / 2 R-r^{2} / w^{2}\right),
\end{aligned}
\end{equation}
where
\begin{equation}\label{6}
\Delta \phi=\sum_{n=0}^{7} H\left(\phi-\frac{\pi}{4} n\right) H\left(\frac{\pi}{4} n+\frac{\pi}{4}-\phi\right) \frac{2 \pi}{8} n,
\end{equation}
and $H(x)$ is the Heaviside function. As $\Delta \phi \approx \phi$, only modes with $(s-t)-(n-m)=1$ contribute most in the $\phi$ integral and the percent of modes is given by $I_{st}=|a_{st}|^2$. Rayleigh range and waist are assumed to be equal to those of incident laser. Our calculations show that $I_{10}\approx79.4\%$, $I_{21}\approx9.9\%$ and $I_{32}\approx3.8\%$, thus, the dominant mode of reflected laser pulse is $\rm{LG_{10}}$ which is consistent with the simulation result in Fig.\ref{fig2}.

 In the third stage, reflected vortex laser collides head-on with energetic electron beam and copious hundreds of $\mathrm{MeV}$ $\gamma$-ray photons are emitted through NCS, see Fig.\ref{fig1}. The QED parameter $\chi_e=e\hbar (m^3_e c^4)|F_{\mu \nu}p^\nu|=(\gamma_e/E_s)\sqrt{(\textbf{E}+\boldsymbol{\beta}\times \textbf{B})^2-(\boldsymbol{\beta}\cdot \textbf{E})^2}$ dominants photon emission probability \cite{QEDparameter1985}, where $F_{\mu \nu}$ is the electromagnetic field tensor, $p^\nu$ is the electron four-momentum and $E_s=(m_e^2c^3)/e\hbar\approx1.3\times10^{18}\rm{Vm^{-1}}$ (corresponding to $I_s\approx4.6\times 10^{29}\rm{W/{cm^2}}$) is the Schwinger electric field, at that intensity electrons can be accelerated to $m_ec^2$ in Compton wavelength distance $\lambda_c=\hbar/m_ec=3.8616\times10^{-13}\rm{m}$ and the vacuum decays to create pairs \cite{Schwinger1951}. $\boldsymbol{\beta}=\textbf{v}/c$ is the normalized velocity of electron, $\gamma_e$ is the electron relativistic factor, and $\textbf{E}$ and $\textbf{B}$ are the electric and magnetic fields acting on electrons. It is worth noting that radiation barely occurs when energetic electron beam co-propagates with intense laser in which $\chi_e\rightarrow0$. While, when electron beam and laser counter-propagate, $\chi_e$ can be described as $\chi_e=2\gamma_e|E_\perp|/E_s$ and the value of which is large and copious $\gamma$-ray photons can be emitted. During the process of NCS, the AM of vortex laser and electron beam are converted to $\gamma$-ray photons, and we will discuss this point later.

\section{RESULTS AND DISCUSSION}
\subsection{Generation of vortex electron beam and $\gamma$-ray}
\begin{figure}[htbp]\suppressfloats
\includegraphics[scale=0.45]{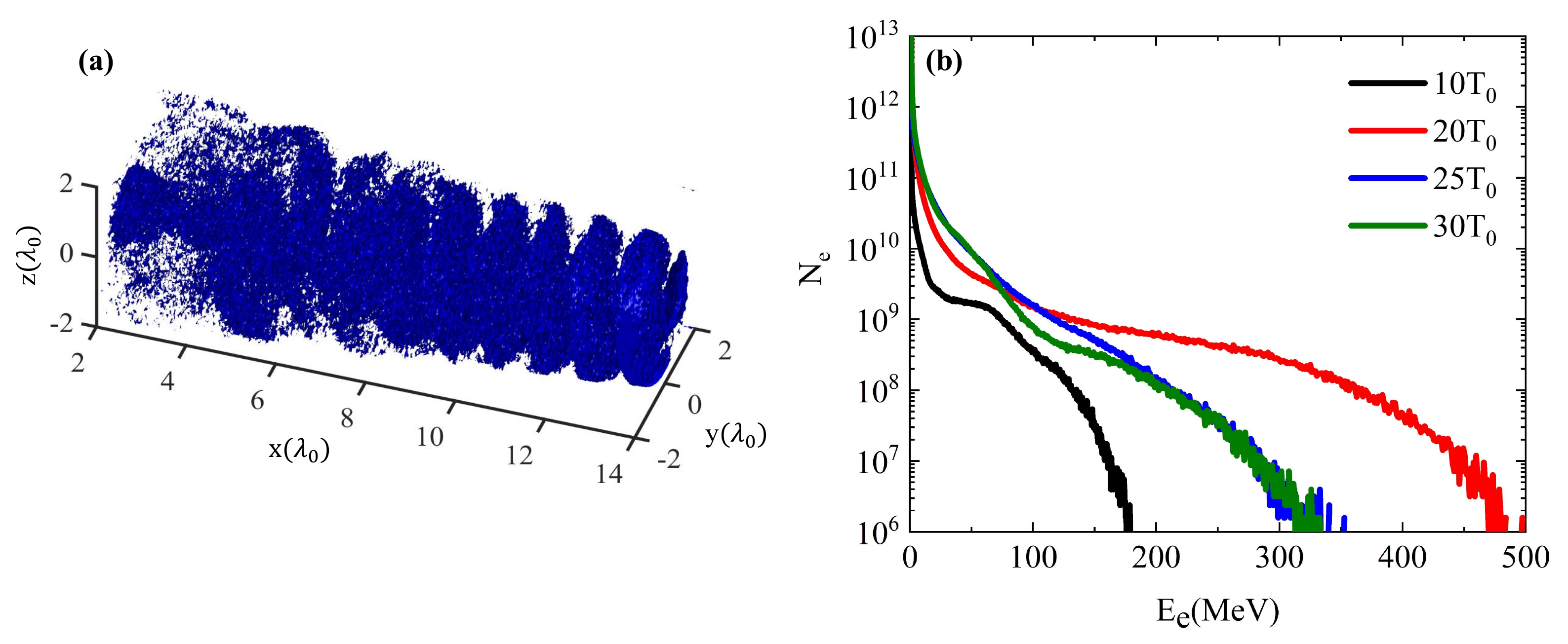}
\caption{\label{fig3}(color online). (a) Three-dimensional isosurface distribution of electron beam at $t=15T_0$. (b) Energy spectrum of electrons at $t=10T_0$, $20T_0$, $25T_0$ and $30T_0$. }
\end{figure}
As indicated in Fig.\ref{fig3}(a), energetic electron beam appears in spiral shape, which results from the helical structure of laser electric field. Figure \ref{fig3}(b) presents the energy spectrum of the electrons at different time. Under the action of the ponderomotive force of laser, the cut-off energy of electrons increases and reaches a maximum of $\sim\rm{500MeV}$ at $20T_0$. And after which, it starts to decrease due to the emission of vortex $\gamma$-ray. The main emission process stopped at about $25T_0$ when CP laser pulse is completely reflected, and the cut-off energy of electrons no longer changes significantly.

\begin{figure}[htbp]\suppressfloats
\includegraphics[scale=0.45]{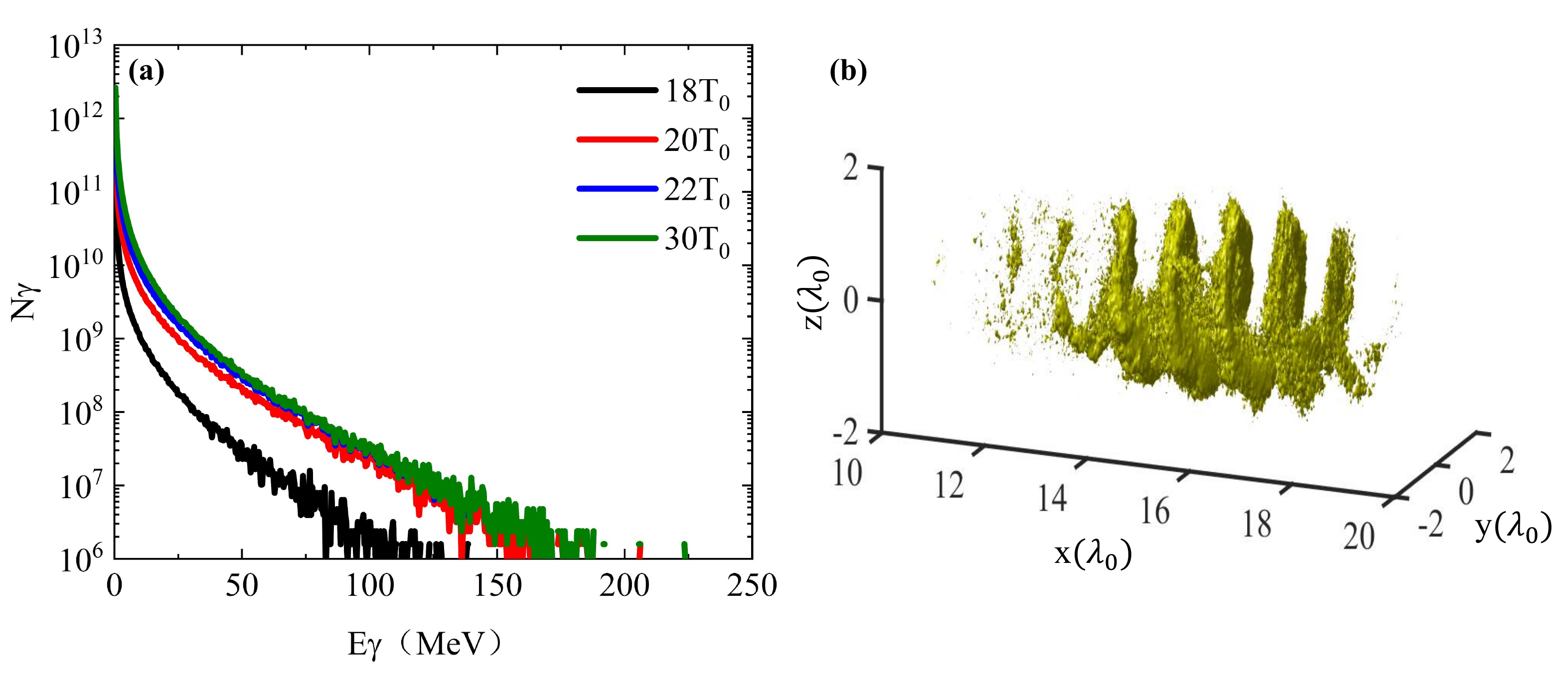}
\caption{\label{fig4}(color online). (a) Energy spectrum of $\gamma$-ray photons at $t=18T_0$, $20T_0$, $22T_0$ and $30T_0$. (b) Three-dimensional isosurface distribution of vortex $\gamma$-ray at $t=20T_0$.}
\end{figure}
$\gamma$-ray photons are emitted via NCS when the reflected vortex laser and energetic electron beam collides head-on. As shown in Fig.\ref{fig4}(a), the cut-off energy of $\gamma$-ray photons increases and reaches a maximum of $\sim\rm{200MeV}$ at about $22T_0$. It is worth noting that the isosurface of vortex $\gamma$-ray, shown in Fig.\ref{fig4}(b), is also helical. Which is because the direction of instantaneously emitted $\gamma$-ray photon is at an angle $1/\gamma_e$ with that of moving electron, and it is negligible for ultrarelativistic electrons ($\gamma_e>>1$). Besides, the $\chi_e$ at $20T_0$ is $0.93$, which coincides with the calculation result by using $\hbar\omega_{\gamma}\approx 0.44\chi_e\gamma m_e c^2$.

\begin{figure}[htbp]\suppressfloats
\includegraphics[scale=0.4]{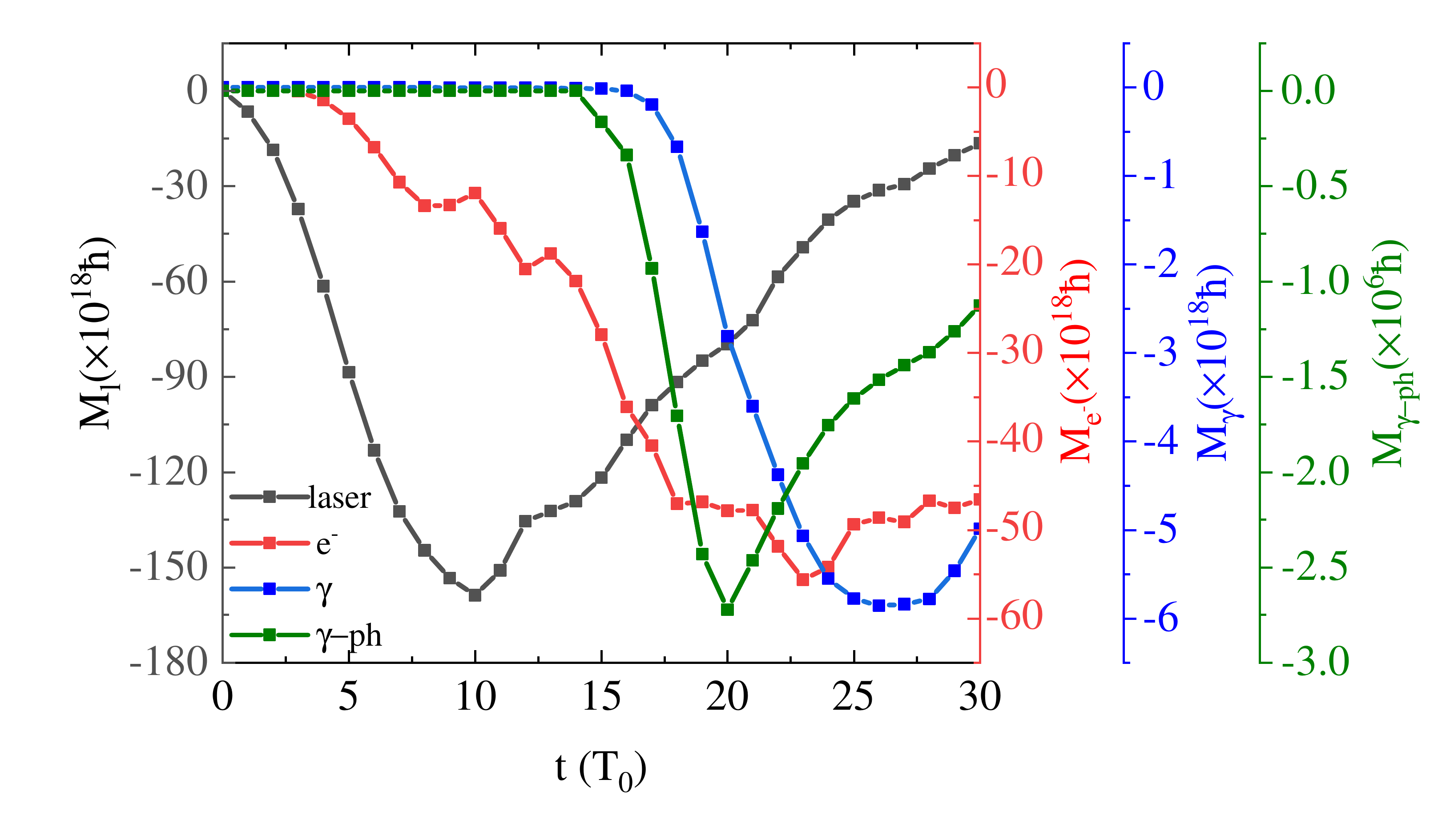}
\caption{\label{fig5}(color online). Evolution of $M_l$, $M_{e^-}$, $M_{\gamma}$ and $M_{\gamma-ph}$ for the AM of laser (black), electron beam (red), vortex $\gamma$-ray (blue) and average of $\gamma$-ray photons (green).}
\end{figure}

In order to figure out the evolution of AM over time, we tracked the different AM, i.e., $M_l$, $M_{e^-}$, $M_{\gamma}$ and $M_{\gamma-ph}$ for the laser, electron beam, vortex $\gamma$-ray and average of $\gamma$-ray photons, as shown in Fig.\ref{fig5}. The AM of Gaussian CP laser can be calculated by
\begin{equation}\label{7}
{M}_l=\big|\varepsilon_0\int\textbf{r}\times(\textbf{E}\times \textbf{B})\rm{d\nu}\big|=\big|\varepsilon_0\triangle \nu\sum_i\textbf{r}_i\times(\textbf{E}_i\times \textbf{B}_i)\big|,
\end{equation}
where $\varepsilon_0$ is the permittivity of vacuum, $\textbf{r}$ is the grid position, $\triangle \nu$ is the volume of one cell and $i$ is the ordinal number of cell. So, the averaged AM of each laser photon is
\begin{equation}\label{8}
M_{l-ph}=\frac{\big|\varepsilon_0\int\textbf{r}\times(\textbf{E}\times \textbf{B})\rm{d\nu}\big|}{\frac{1}{2}\int(\varepsilon_0\textbf{E}^2+\frac{1}{\mu_0}\textbf{B}^2)\rm{d\nu}}\hbar\omega_0
=(l+\delta)\hbar,
\end{equation}
where $\frac{1}{2}\int(\varepsilon_0\textbf{E}^2+\frac{1}{\mu_0}\textbf{B}^2)\rm{d\nu}$ is the total electromagnetic energy of laser, $\mu_0$ is the vacuum permeability. The black line in Fig.\ref{fig5} shows that the $M_l$ reaches the maximum of $1.58\times 10^{20}\hbar$ which is close to that in the vacuum $2\times 10^{20}\hbar$ at $10T_0$ when the CP laser completely enters the simulation region, and the difference is transferred to the electrons in cone.

 The AM of particle beam with respect to the $x$-axis is
\begin{equation}\label{9}
M_{particle}=\big|\sum_j(\textbf{r}_j\times \textbf{p}_j)\big|=\big|\sum_j(yp_z-zp_y)_j\hat{\textbf{x}} \big|,
\end{equation}
where $\textbf{r}_j$ and $\textbf{p}_j$ are the position and momentum of the $j$-th particle respectively, and $p_y$ and $p_z$ are the momentum in the $y$ and $z$ direction. As shown by the red line in Fig.\ref{fig5}, the $M_{e^-}$ increases since CP laser propagates into the cone-channel and reaches maximum of $5.56\times10^{19}\hbar$ at $23T_0$ and then decreases. In the first stage, before $15T_0$, electrons are accelerated by CP laser and the SAM of laser is converted to the OAM of electrons. In the second stage, the electrons are still being accelerated because the intensity of the forward laser is greater than that of the reflected laser, thus, $M_{e^-}$  still increases. In the third stage, due to  radiation reaction, the AM of the reflected laser can both be transferred to the electrons and $\gamma$-ray photons \cite{16}. Therefore, the $M_{e^-}$ continues to increase. After the third stage, i.e., $23T_0$, there are no sources of OAM for electrons, and $M_{e^-}$ decreases gradually.
\begin{figure}[ht]\suppressfloats
\includegraphics[scale=0.50]{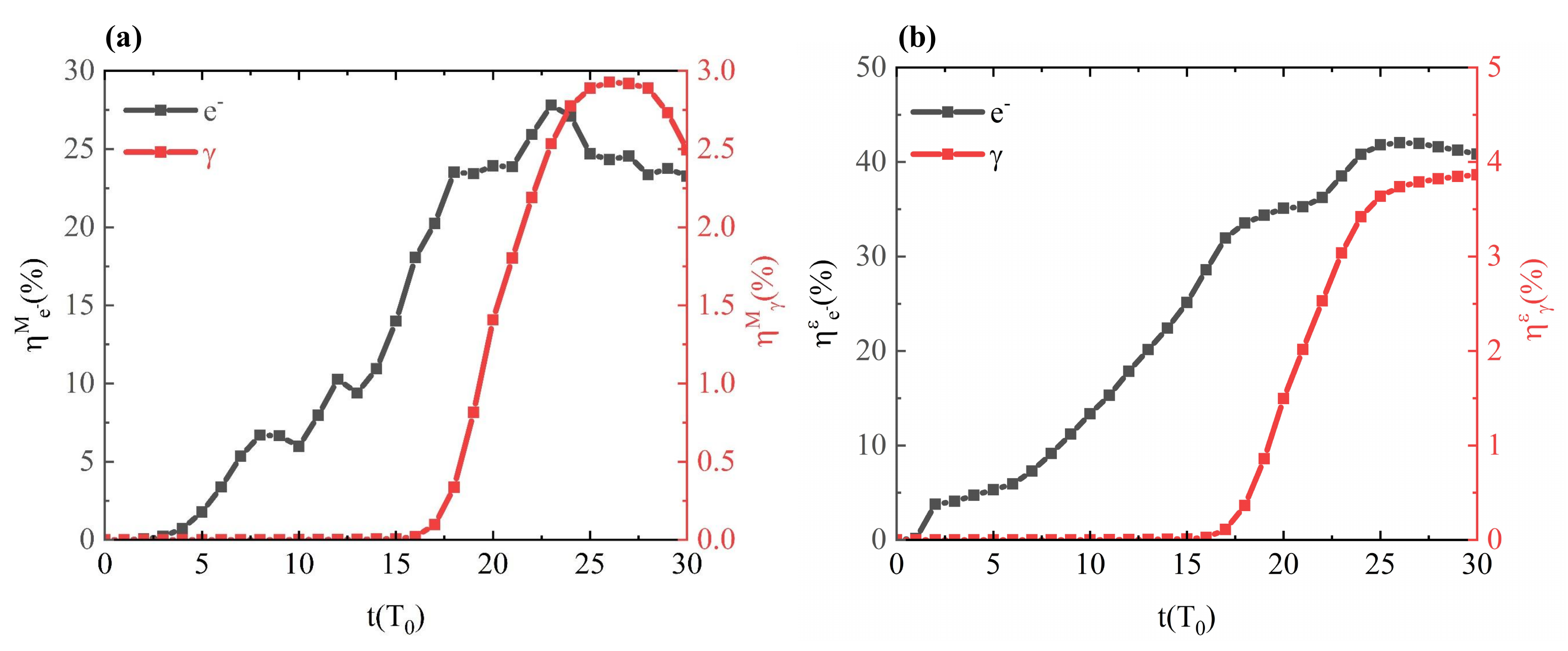}
\caption{\label{fig6}(color online). (a) Evolution of conversion efficiency of laser AM for electron beam $\eta^M_{e^-}$ (black) and vortex $\gamma$-ray $\eta^M_{\gamma}$ (red). (b) Evolution of conversion efficiency of laser energy for electron beam $\eta^{\varepsilon}_{e^-}$ (black) and vortex $\gamma$-ray $\eta^{\varepsilon}_{\gamma}$ (red).}
\end{figure}

As shown by the blue line of $M_{\gamma}$ and the green line of $M_{\gamma -ph}$ in Fig.\ref{fig5}, both of them increase since $t=15T_0$ when NCS occurs and reach maximums of $5.9\times 10^{18}\hbar$ at $26T_0$ and $2.8\times10^6\hbar$ at $20T_0$, respectively. And the OAM of $\gamma$-ray photons comes from that of the reflected laser. It is worth noting that $M_{\gamma}$ and $M_{\gamma -ph}$ reach maximums at different time, this is because there are always $\gamma$-ray photons being emitted after $15T_0$ and the number of $\gamma$-ray photons at $26T_0$ is more than that at $20T_0$.
\begin{figure}[ht]\suppressfloats
\includegraphics[scale=0.50]{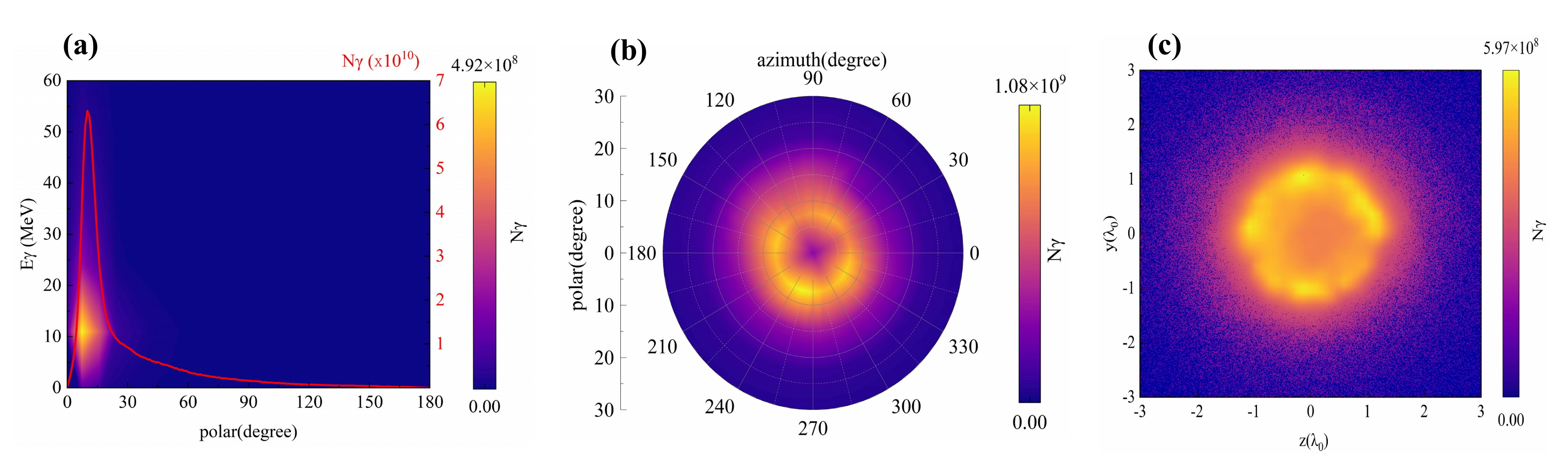}
\caption{\label{fig7}(color online). (a) The polar-energy distribution of $\gamma$-ray photons at $t=20T_0$. The red line exhibits polar-number distribution. (b) The angular distribution of $\gamma$-ray photons at $t=20T_0$. (c) The position distribution of $\gamma$-ray photons at $t=20T_0$.}
\end{figure}

The conversion efficiency of laser AM for particles can be calculated by
\begin{equation}\label{10}
\eta^M_{particle}=\frac{|\sum_j(\textbf{r}_j\times \textbf{p}_j)|}{|\varepsilon_0\triangle \nu\sum_i\textbf{r}_i\times(\textbf{E}_i\times \textbf{B}_i)|}.
\end{equation}
So, the conversion efficiencies of laser AM to electron beam $\eta^M_{e^-}$ and vortex $\gamma$-ray $\eta^M_{\gamma}$ are roughly $27.8\%$ and $3\%$, as shown in Fig.\ref{fig6}(a), and $\eta^M_{\gamma}$ is almost 4 times higher than that of the channel scheme in Ref.\cite{channel2021HPL} and 1.5 times higher than that of Ref.\cite{channel2022PST}. Furthermore, at the end of the simulation, the laser energy conversion efficiencies to electron beam  $\eta^{\varepsilon}_{e^-}$ and vortex $\gamma$-ray $\eta^{\varepsilon}_{\gamma}$ are around $41\%$ and $3.8\%$, see Fig.\ref{fig6}(b), which are double those in Ref.\cite{channel2021HPL}.

The polar angle of particle can be calculated by $\arcsin(\sqrt{p^2_\bot/p^2})$ for $p_x\geq0$ and $\pi-\arcsin(\sqrt{p^2_\bot/p^2})$ for $p_x<0$, where $p^2_\bot=p^2_y+p^2_z$ and $p^2=p^2_\bot+p^2_x$. Figure \ref{fig7}(a) indicates that $\gamma$-ray photons are concentrated in a narrow range of $5^{\circ}$ centered around polar angle $7^{\circ}$, the corresponding photon energy is $10\rm{MeV}$ and the number of $\gamma$-ray photons is about $1\times10^{9}$, which are all rather better. Figures \ref{fig7}(b) and \ref{fig7}(c) show that the distribution of  momentum and position of $\gamma$-ray photons in $y-z$ plane are in the shape of a doughnut, which is consistent with the helical structure of the vortex $\gamma$-ray. From Fig.\ref{fig7}(c), it can be roughly estimated that $\gamma$-ray photons are distributed in the radial range $1\lambda_0$-$1.2\lambda_0$. And as can be seen from Fig.\ref{fig4}(b), the duration of vortex $\gamma$-ray pulse $\tau_1\approx6\lambda_0/c=20\rm{fs}$. Thus, the peak brilliance of vortex $\gamma$-ray is $\sim5\times10^{22}$ photons ${\rm\cdot s^{-1} \cdot mm^{-2} \cdot mrad^{-2}}$ $0.1\%$ bandwidth at $10\rm{MeV}$, which is promising for some potential applications in various domains.
\subsection{Vortex $\gamma$-ray emission of different schemes}
\begin{figure}[htbp]\suppressfloats
\includegraphics[scale=0.4]{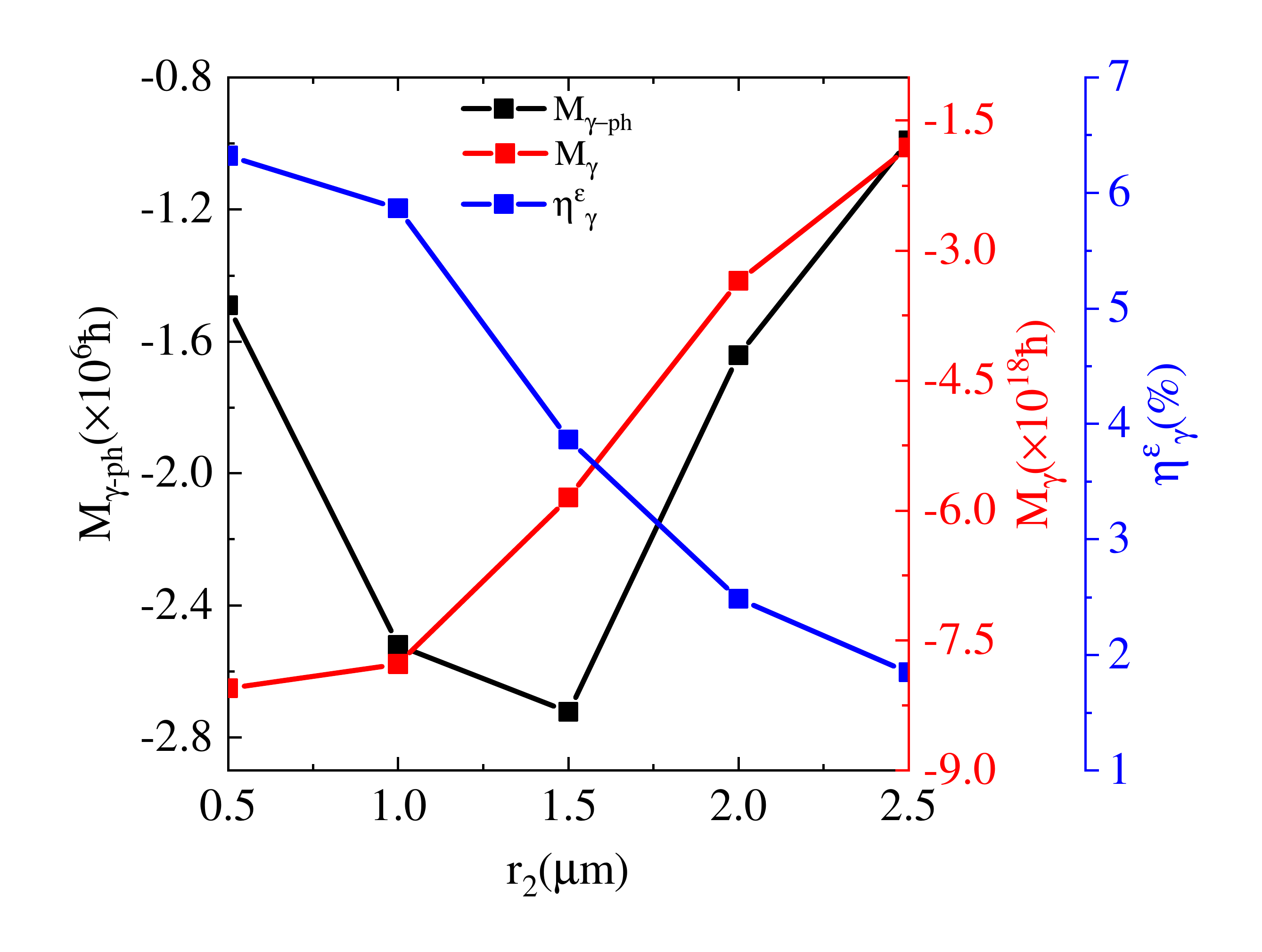}
\caption{\label{fig8}(color online). The evolution of $M_{\gamma}$ (red), $\eta^{\varepsilon}_{\gamma}$ (blue) and $M_{\gamma-ph}$ (black) with the increase of right radius $r_2$. }
\end{figure}

\begin{table}[htbp]
 \setlength\tabcolsep{3pt}
    \caption{\label{tab1}The maximums of $M_l$, $M_{e^-}$, $M_{\gamma}$, $M_{\gamma-ph}$, $\eta^M_{e^-}$ and $\eta^M_{\gamma}$ of three cases of CP-cone-fan, LG-cone-fan and LG-cone-foil.}
    \begin{center}
        \begin{tabular*}{15cm}{@{\extracolsep{\fill}}l c c c c c c c c c c}
            \hline\hline
            & case & $M_l(\times 10^{18}\hbar)$ & $M_{e^-}(\times 10^{18}\hbar)$ & $M_{\gamma}(\times 10^{18}\hbar)$ & $M_{\gamma-ph}(\times 10^{6}\hbar)$ & $\eta^M_{e^-}$ & $\eta^M_{\gamma}$
            \\ \hline
            &CP-cone-fan & $-200$ & $-55.6$ & $-5.9$ & $-2.8$ & $27.8$ & $2.9$ \\ \hline
            &LG-cone-fan & $-400$ & $-92.9$ & $-2.0$ & $-1.1$ & $23.2$ & $0.5$   \\ \hline
            &LG-cone-foil& $-400$ & $-98.6$ & $-2.1$ & $-0.9$ & $24.7$ & $0.5$
            \\ \hline \hline
        \end{tabular*}
    \end{center}
\end{table}
\begin{figure}[htbp]\suppressfloats
\includegraphics[scale=0.65]{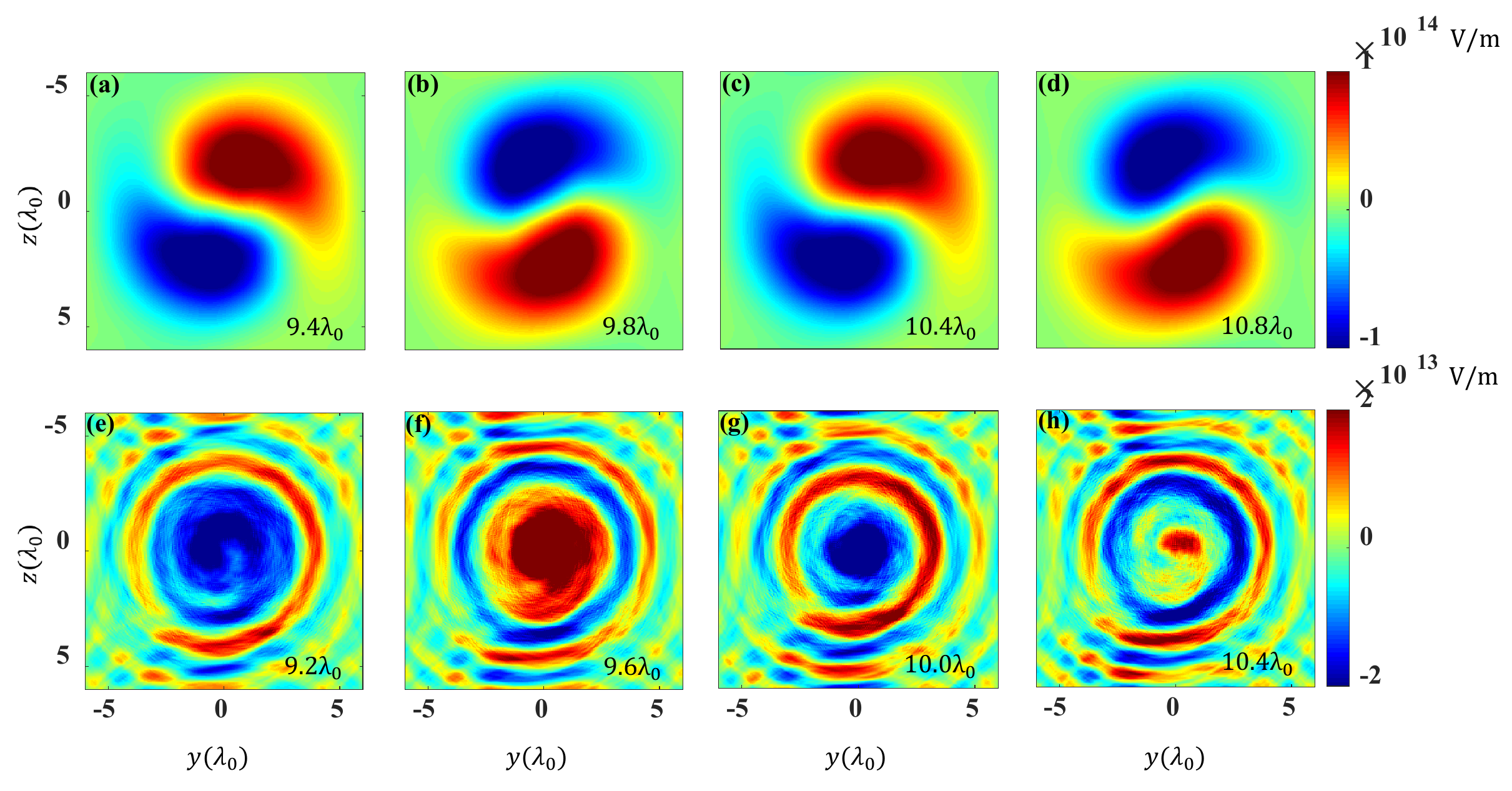}
\caption{\label{fig9}(color online). (a)-(d): Distributions of transverse electric field $E_y$ of incident $\rm{LG_{10}}$ laser at different crossing sections at $t=15T_0$. (e)-(h): Distributions of transverse electric field $E_y$ at different crossing sections at $t=25T_0$ when the incident laser is completely reflected by fan.  }
\end{figure}
We have performed comparative simulations for several different right radius $r_2$ while keeping other parameters the same. Figure \ref{fig8} indicates that the red line of $M_{\gamma}$ and the blue line of $\eta^{\varepsilon}_{\gamma}$ all decrease with the increase of $r_2$. As $r_2$ increases, the cone becomes less constrained to the incident laser and fewer electrons are dragged out of the cone. Therefore, the average energy and density of the electrons decrease \cite{cone2016JAP}. And accordingly, the $M_{\gamma}$ and $\eta^{\varepsilon}_{\gamma}$ decrease. However, $M_{\gamma-ph}$, the black line in Fig.\ref{fig8}, reaches maximum of $\sim2.8\times10^6\hbar$ at $r_2=1.5\rm{\mu m}$ which indicates that $r_2=1.5\rm{\mu m}$ is the best choice when we want the highest AM of $\gamma$-ray photons.

We also studied the case of CP-LG laser, which remained the same laser energy as the CP one. The LG laser is left-hand and in $\rm{LG_{10}}$ mode, and the target attached behind are fan and foil respectively, i.e., LG-cone-fan and LG-cone-foil cases. The equation of $\rm{LG_{10}}$ laser is

\begin{equation}\label{11}
\begin{aligned}
\mathrm{LG}_{10}=&(C_{10}/w)\exp(-ikr^2/2R)\exp \left({-r^{2}}/{w^{2}}\right)\\
&\times \exp(-i\psi)\exp (-i\phi)\\
&\times \left({\sqrt{2} r}/{w}\right)^{1} L_{0}^{1}\left({2 r^{2}}/{w^{2}}\right).
\end{aligned}
\end{equation}
So, equation \ref{5} becomes
\begin{equation}\label{12}
\begin{aligned}
a_{s t}=&\left\langle\mathrm{LG}_{s t}|\exp (-i \Delta \phi)| \mathrm{LG}_{nm}\right\rangle \\
=& \iint r d r d \phi\left(C_{s t}^{*} / w_{s t}\right) \exp \left(i k r^{2} / 2 R_{s t}-r^{2} / w_{s t}^{2}\right) \\
& \times \exp [-i(s-t) \phi](-1)^{\min (s, t)}\left(r \sqrt{2} / w_{s t}\right)^{|s-t|} \\
& \times L_{\min (s, t)}^{|s-t|}\left(2 r^{2} / w_{s t}^{2}\right) \exp (-i \Delta \phi) \\
& \times\left(C_{10} / w\right) \exp \left(-i k r^{2} / 2 R-r^{2} / w^{2}\right)\\
& \times \exp (-i\phi)\left(r \sqrt{2} / w\right)^{1}\times L_{0}^{1}\left(2 r^{2} / w^{2}\right),
\end{aligned}
\end{equation}
where $L_{0}^{1}\left(\frac{2 r^{2}}{w^{2}}\right)=1$. The photon of $\rm{LG_{10}}$ laser carries both OAM and SAM of $(l+\delta)\hbar=-2\hbar$, which is twice that of CP laser, see Table \ref{tab1}. Consequently, the $M_{e^-}$ and $\eta^M_{e^-}$ in both cases are almost twice that in CP case. During reflection, modes with $(s-t)-(n-m)=1$ contribute most, and our calculations show that $I_{20}\approx89\%$ and $I_{31}\approx7.4\%$. While, $M_{\gamma}$, $M_{\gamma-ph}$ and $\eta^M_{\gamma}$ are all lower than those in CP case which is because that the transverse electric field $E_y$ of the reflected laser is much lower than that of incident $\rm{LG_{10}}$ laser, as shown in Fig.\ref{fig9}. And the different is lost during the reflection process. Therefore, the CP laser is much more appropriate than LG laser in our scheme. Additionally, comparative simulations also indicate that the AM of vortex $\gamma$-ray is influenced by the mode of laser and the number of steps of fan.

\section{CONCLUSION}
In conclusion, we have proposed an all-optical scheme to generate a bright collimated vortex $\gamma$-ray with large OAM by simulating a CP laser irradiates on a cone-fan target. In contrast to other schemes for generating vortex $\gamma$-rays, the SAM of CP laser is converted into the OAM of electron beam and vortex $\gamma$-ray with high efficiencies of $27.8\%$ and $3\%$, the energy conversion efficiency of laser for electron beam and vortex $\gamma$-ray are around $41\%$ and $3.8\%$, and the OAM of vortex $\gamma$-ray and averaged $\gamma$-ray photons are $6\times 10^{18}\hbar$ and $3\times 10^{6}\hbar$. The peak brilliance of vortex $\gamma$-ray is about $5\times10^{22}$ photons ${\rm\cdot s^{-1} \cdot mm^{-2} \cdot mrad^{-2}}$ $0.1\%$ bandwidth at $10\rm{MeV}$ and divergence angle $5^{\circ}$. The properties of this bright collimated vortex $\gamma$-ray might enable the development of novel application in materials and laboratory astrophysics.

Comparative simulations of different right radius $r_2$ show that $M_{\gamma-ph}$ reaches maximum at $r_2=1.5\rm{\mu m}$, and comparative simulations of different laser modes indicate that CP laser is better than LG laser in our scheme. In addition, it can be inferred from our study that both the mode of laser and the number of steps of fan can also affect the generation of vortex $\gamma$-rays, which is worthy of further study.

\begin{acknowledgments}
 This work was supported by the National Natural Science Foundation of China (NSFC) under Grant No.11875007 and No. 11935008. The computation was carried out at the HSCC of the Beijing Normal University. We acknowledgment the open source PIC code EPOCH.
\end{acknowledgments}

\end{document}